\documentclass[twocolumn,3p,times]{elsarticle}

\usepackage{graphics} 
\usepackage{graphicx} 
\usepackage{epstopdf}
\usepackage{color}
\usepackage{scrtime}
\usepackage{amssymb}
\usepackage{amsmath}
\usepackage{amsfonts}
\usepackage{natbib}
\usepackage{physics}
\usepackage{booktabs} % For professional looking tables
\usepackage{upgreek}
\usepackage{hyperref}

\usepackage[final]{changes}
\journal{Physics Letters B}

\begin{document}

\begin{frontmatter}

\title{ Weisskopf units for neutron-proton pair transfers}

\author[1,2]{J.~A.~Lay}
\author[3,4]{Y. Ayyad}
%\fnref{myfootnote1}}
\author[5]{A.~O. Macchiavelli\corref{mycorrespondingauthor}}
\cortext[mycorrespondingauthor]{Corresponding author}
\ead{aom@lbl.gov}
%\fntext[myfootnote1]{Present address:  University of Santiago ....}

 \address[1]{Departamento de F\'isica At\'omica, Molecular y Nuclear, Facultad de F\'isica, Apartado 1065,
E-41080 Sevilla, Spain.}

\address[2]{Instituto Interuniversitario Carlos I de Física Teórica y Computacional (iC1), E-41080 Sevilla, Spain.}

\address[3]{Facility for Rare Isotope Beams, Michigan State University East Lansing, MI 48824, USA.}
\address[4]{IGFAE, Universidade de Santiago de Compostela, E-15782, Santiago de Compostela, Spain.}   
  \address[5]{  Nuclear Science Division, Lawrence Berkeley National Laboratory, Berkeley, California 94720, USA.}

\date{\today}

\begin{abstract}
We introduce the concept of neutron-proton two-particle units ($np$-Weisskopf units) to be used in the analysis of the ($^3$He,$p)$ and $(p,^3$He) \added{reactions on nuclei} along the N=Z line.  These are presented for the conditions relevant to the $(n,j,\ell$) orbits expected from $^{16}$O to $^{100}$Sn.  As is the case of the Weisskopf units for electromagnetic transitions,  the $np$-WU's will provide a simple, yet robust, measure of isoscalar and isovector  $np$ pairing \replaced{collective effects}{collectivity}. 
\end{abstract}

\end{frontmatter}

%\linenumbers

\section{Introduction}

In 1958, Bohr, Mottelson and Pines \cite{bmp} suggested a pairing mechanism in the atomic nucleus 
analogous to that observed in superconductors~\cite{BCS}.  Since the publication of that seminal paper, 
 a wealth of experimental data have been accumulated, supporting the important role played by 
 neutron-neutron and proton-proton ``Cooper pairs'' in modifying many nuclear properties such as 
 deformation, moments of inertia, alignments, etc.~\cite{BrogliaBrink,50yearsBCS,DeanMorten}. Driven by advances in experimental techniques, the development of sensitive and highly 
efficient instruments  and the availability of radioactive beams, the study of pairing correlations in exotic nuclei 
is a subject of active research in nuclear physics. Of particular interest is the competition between isovector and isoscalar  ``Cooper pairs" expected to occur in N $\approx$ Z nuclei~\cite{review}.

The dominant pairing in almost all known nuclei with N $>$ Z is that in which ``superconducting” pairs of neutrons ({\sl nn}) and protons ({\sl pp}) couple to a state with angular momentum $J=0$ and isospin $\tau=1$, known as isovector or spin-singlet pairing.  However, for nuclei with N $\approx$ Z, neutrons and protons occupy the same single-particle orbits at their respective Fermi surfaces and pairs, consisting of a neutron and a proton ({\sl np}), may form. These types of pairs couple in either isovector or isoscalar (spin-triplet with angular momentum $J=1$ and isospin $\tau=0$) modes, the latter being allowed by the Pauli principle. Since the nuclear force is charge independent, we expect to observe the effects of
the standard $\tau=1$ pairing on an equal footing between the $\tau_z=0$ ($np$) and 
$|\tau_z|=1$ ($nn$ and $pp$) components.  Furthermore, given that the nuclear force is stronger in the $T=0$ channel, {\sl a priori}  arguments suggest the existence of correlated isoscalar $np$ pairs. However, the effectiveness of the in-medium $T=0$ correlations in giving rise to a ``deuteron-like condensate"  remains a controversial and fascinating topic in nuclear structure physics~\cite{review,aom}.       	
	
\section{The Experimental probe}

Two-neutron transfer reactions such as $(p,t)$ and $(t,p)$ have provided a unique tool to understand neutron pairing correlations in nuclei~\cite{Yos62,Bayman68,Broglia73}. Based on the formal analogy between pairing distortions and quadrupole shape fluctuation~\cite{Broglia73,BM,BTG}, where an important measure of collective effects is provided by the reduced transition probabilities (i.e. $B(E2)$'s), one can associate a similar role to the transition operators $\langle f|a^\dagger a^\dagger|i\rangle$ and $\langle f|a a|i\rangle$ in the two-particle transfer mechanism between the initial $|i\rangle$ and
final $|f\rangle$ states.

\begin{figure}
\centering\includegraphics[trim=0 120 260 180,clip,width=6.0cm,angle=90]{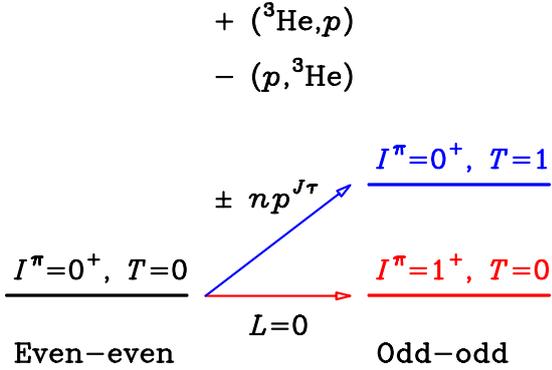}
\caption{\label{fig1} (Color online) Schematic of the reactions.}
\end{figure}

Thus, it seems natural to consider the transfer of an $np$ pair from even-even to odd-odd self-conjugate nuclei as a sensitive probe to study $np$ correlations~\cite{Fro71,Isa05,Alex18}.  Of the possible direct reactions we could envision, the $(^3$He,$p)$ and $(p,^3$He) are perhaps the best choice since both isoscalar and isovector transfers are allowed. 
As schematically showed in Fig.~\ref{fig1}, exclusive forward center of mass angles ($L=0$) cross sections, $d\sigma^{IT}/d\Omega(\theta)$ populating the lowest $I^\pi(T) = 0^+(1), 1^+(0)$ states in the final odd-odd N=Z nuclei could provide a robust observable to quantify the nature of and interplay between spin-triplet and spin-singlet superfluidity.  Specifically, the ratio
$\mathcal{R}_{01} = d\sigma^{01}/d\sigma^{10}$ is appealing since experimental systematic uncertainties and, to some extent, kinematic conditions of the reaction cancel out.   

The ultimate goal for these studies is to explore the region of superfluid nuclei between $^{56}$Ni and $^{100}$Sn. Initial measurements of these reactions with radioactive beams in reverse kinematics done at ATLAS/ANL~\cite{aom2} and at GANIL~\cite{marlene}, showed the feasibility of such a program which will be featured prominently at new rare isotopes accelerator facilities. 
Recently, systematic measurements of these reactions in N=Z $sd$-shell nuclei were carried out at RCNP~\cite{Ayy17} under the same experimental conditions. The experimental data together with a detailed DWBA analysis established a valuable baseline for further studies beyond $^{40}$Ca.

\section{The $np$ two-particle units}

The question we would like to address in this work is how to (empirically) assess, from the measured cross-sections and/or $\mathcal{R}_{01}$, the collective effects due to pairing correlations?   

The single-particle units (WU) introduced by Weisskopf~\cite{WU} for electromagnetic transitions offer an effective way to gauge collectivity in shape degrees of freedom. The famous example of measured $B(E2)$'s in WU clearly reveals the regions of the nuclear chart where quadrupole deformations develop~\cite{BM}; in these regions the measured transition rates could be up to several hundred's WU.  In Ref.~\cite{Broglia72} two-particle units relevant for two-neutron transfers in $(t,p)$ and $(p,t)$ reactions were discussed.  Here, in line with the above, we introduce the $np$ two-particle units to be used in the analysis of the $(^3$He,$p)$ and $(p,^3$He) reactions.

Let us start by recalling that in superfluid nuclei, where the BCS theory provides a good representation of the ground states, the cross-section for two-neutron transfers from the nucleus $A_0$ to $A_0\pm 2$ is given approximately by~\cite{Yos62}:
\begin{equation}
d\sigma/d\Omega \approx \lvert \sum_j U_j V_j \rvert ^2 (d\sigma/d\Omega)_{2sp}= (\frac{\Delta}{G})^2 (d\sigma/d\Omega)_{2sp}
\label{eq1}
\end{equation}
where $U_j$ and $V_j$ are the probability amplitudes for the orbit $j$ to be empty
and occupied respectively,  $\Delta$ is the pairing gap and $G$ is the strength of the pairing interaction.  Eq. (1) explicitly shows the enhancement due to the coherent contributions of the correlated $nn$ pairs. With typical values of $\Delta \sim 12/\sqrt{A}$ MeV and $G \sim 20/A$ MeV, the enhancement factor is $\sim A/4$, obviously increasing with $A$ as expected from the larger number of available orbits for the pairs to scatter into.

For the case at hand, we look at the experimental ratio $\mathcal{R}_{01}$  %\sigma(I^\pi=0^+)/\sigma(I^\pi=1^+)$ 
in terms of two-particle units:
\begin{equation}
\frac{\mathcal{R}_{01}}{\mathcal{R}_{01,2sp}} =  \frac{d\sigma^{01}/d\sigma^{01}_{2sp}}{d\sigma^{10}/d\sigma^{10}_{2sp}}
\label{eq2}
\end{equation}
as a relative measure to signal possible enhancement effects between the two isospin channels.   For a single $np$ pair transfer, the cross-section factorizes in a structure part, $\mathcal{S}$, and a DWBA reaction part usually calculated with codes such as DWUCK~\cite{Kuhn} or FRESCO~\cite{Ian}, 
\begin{equation}
d\sigma/d\Omega_{2sp}= \mathcal{S} \sigma_{DW}
\label{eq3}
\end{equation}

Let us assume that the $np^{\tau J}$  pairs are constructed from a given single-particle orbit with ($n \ell j$) orbit quantum numbers,  appropriate for the case under study, and that A and B are respectively the initial and final nuclei.  Since we start from an even-even nucleus with $I_A=0$ and $T_A=0$, it follows from the selection rules that $I_B=J$ and $T_B=\tau$, then the structure factors are:
%%$\mathcal{S}(I)$, assuming . 

\begin{equation}
\mathcal{S}^{\pm}(I)= %\mathcal{N}  
(C^{J\tau}_{^3He,p})^2 S^{J\tau}_{^3He,p}(C^{J\tau}_{A,B})^2 S^{J\tau}_{A,B}
\label{eq4}
\end{equation}
where 
\begin{equation}
  C^{J\tau}_{^3He,p}=\langle 1/2 -1/2 \tau 0 | 1/2 -1/2 \rangle \nonumber
\end{equation}

and depending on the case for:

\begin{itemize}
    \item Pair addition  A$(^3$He,$p)$B
    
    \begin{equation}
  C^{J\tau}_{A,B}=\langle 0 0  \tau 0 | \tau  0 \rangle \nonumber
\end{equation}
%\begin{equation}
%\mathcal{S}^+(I)= %\mathcal{N}  
%(C^{J\tau}_{^3He,p})^2 S^{J\tau}_{^3He,p}(C^{J\tau}_{A,B})^2 S^{J\tau}_{A,B}
%\label{eq4}
%\end{equation}
  %\frac{(2I_B+1)}{(2I_A+1)(2J+1)}  
\item Pair removal  A$(p,^3$He)B
%\begin{equation}
%\mathcal{S}^-(I)= \frac{1}{(2J+1)}\mathcal{S}^+(I)
%\label{eq5}
%\end{equation}

\begin{equation}
  C^{J\tau}_{A,B}=\langle \tau  0  \tau 0 | 0 0  \rangle \nonumber
\end{equation}

\end{itemize}
%\frac{(2I_A+1)}{(2I_B+1)}
are the corresponding isospin projection Clebsch-Gordan coefficients. Within the context of two-particle units, we further assume transfers of 2 nucleons in a $j^2$ configuration to a given core, and the spectroscopic amplitudes used are $S^{J\tau}_{A,B}=1$ with a center of mass correction equal to $\left((A+2)/A\right)^{2n+\ell}$ as usually needed in Shell Model Calculations~\cite{Any74}.  In Table~\ref{Tab1} we give the common factors entering in Eq.~(\ref{eq4}).% and \ref{eq5}.
%\begin{equation}
%S^{J\tau}_{A,B}=\sqrt{\frac{n(n-1)}{2}}|(n-2)(0,0);2(J,\tau)|\}n(J,\tau)\rangle={n \choose 2} c.f.p
%\label{eq5}
%\end{equation}

\begin{figure}
\centering\includegraphics[trim=60 190 60 160,clip,width=8.5cm,angle=0]{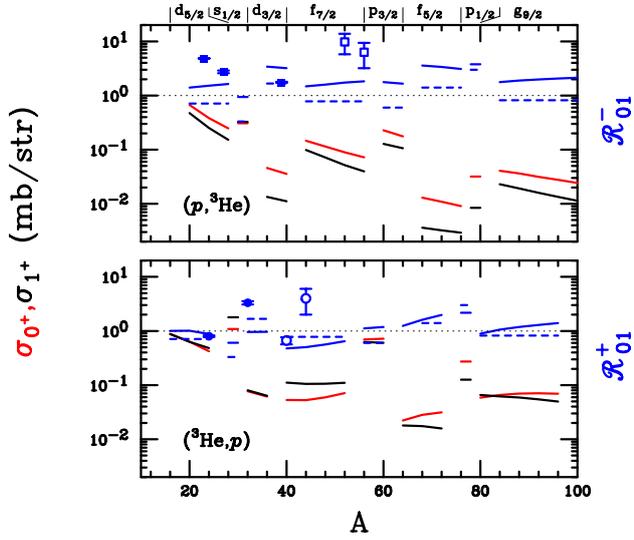}
\vspace{0.1cm}
\caption{(Color online) Left axis: Unit cross-sections for the $0^+$ (red) and $1^+$ (black) states and Right axis: $np$ Weisskopf Units (blue) as a function of the target mass. The results from the approximation of Eq.~\ref{eq8} is shown in a dashed (blue) line. The corresponding orbits being filled are indicated at the top. Experimental data points from Refs.~\cite{aom2,marlene,Ayy17} are also included.  }
\label{fig2}
\end{figure}

Being interested only in $L=0$ transfers we consider the limit $\theta \rightarrow 0$ for the DW cross-sections.  It follows from the Eqs. above that

\begin{equation}
\mathcal{R}_{01,2sp}^{\pm} = \frac{ \mathcal{S}^{\pm}(0^+)}{\mathcal{S}^{\pm}(1^+)} \frac{\sigma_{DW}^{n \ell j, 01}}{\sigma_{DW}^{n \ell j, 10}},
\label{eq6}
\end{equation}
which we introduce as the $np$ two-particle units (or $np$-Weisskopf units).  The unit cross-sections and the WU's as a function of the target mass A are shown in Figs.~\ref{fig2}. The calculations were performed in second-order DWBA~\cite{Tho13} with the code FRESCO~\cite{Ian} at 25~MeV for the $(^3$He,$p)$ reaction and 65 MeV/A 
for the $(p,^3$He) reaction, with conditions relevant to the filling of the 
different $(n,j,\ell$) orbits at the N=Z line, from $^{16}$O to $^{100}$Sn. Optical potentials between the counterparts of the reactions were chosen to be CH89 for protons~\cite{CH89}, Daehnick
for deuteron potentials~\cite{Dae80}, and Bechetti-Greenlees for the $^{3}$He potentials~\cite{ADNTD}.\added{ For the calculation of these potentials we have used the code FR2IN~\cite{Brown}.} A final important ingredient for the kinematics is the Q-value which depends on the masses and the energy of the final 0$^+$ and 1$^+$ states. For this purpose we have used the experimental value when available or that from the systematics otherwise~\cite{review,AME2020}.

\begin{table}
  \caption{\label{Tab1}}
  \bigskip
  \begin{tabular*}{\columnwidth}{c c c c c}
    \hline
    Reaction& $(I,T)$ &~ $(C^{J\tau}_{^3He,p})^2$ & $S^{J\tau}_{^3He,p}$ & $(C^{J\tau}_{A,B})^2$ \\%& $2J+1$ \\ 
 \hline\hline
 $(^3$He,$p)$ &(0,1)  & $\frac{1}{3}$& $\frac{3}{2}$ & 1 \\ [2pt] %& -   \\[2pt]
   &(1,0)  & 1 & $\frac{3}{2}$ & 1 \\ [2pt] %& - \\[2pt]
 \hline
  $(p,^3$He)  &(0,1) &  $\frac{1}{3}$ & $\frac{3}{2}$ & $\frac{1}{3}$ \\ [2pt] %& 1 \\[2pt]
  &(1,0) & 1  & $\frac{3}{2}$ & 1 \\ [2pt] %& 3  \\[2pt]
  \hline
  \end{tabular*}
\end{table}

A simple analytical expression for the $np$-WU's can be obtained from the following approximation.  Since the energy difference of the $T=0$ and $T=1$ low-lying states in the odd-odd final nucleus is small compared to the reaction and binding energy scales, it is expected that the reaction kinematics part will cancel out,  leaving only the different probabilities of finding in the $(n \ell j)^2$ configuration, an $np$ pair in relative $^3S_1$ or $^1S_0$ states entering in the pair form factor.

Following~\cite{Norman}, the cross sections for stripping ($^3$He,$p$) or pick-up ($p$,$^3$He) can be reduced to the same formula:

\begin{eqnarray}
\left. \frac{d\sigma}{d\Omega} \right|_{\rm stripping} & = & \frac{k_f}{k_i}\frac{2I_B+1}{2I_A+1}\left.\frac{d\sigma}{d\Omega}\right|_{0} \\
\left. \frac{d\sigma}{d\Omega} \right|_{\rm pick-up} & = & \frac{k_f}{k_i}\frac{2j_p+1}{2j_{^3{\rm He}}+1}\left.\frac{d\sigma}{d\Omega}\right|_{0}
\end{eqnarray}
where:

\begin{equation}
    \left.\frac{d\sigma}{d\Omega}\right|_{0}=\frac{\mu_{p}\mu_{^3{\rm He}}}{(2\pi\hbar^2)^2}\sum_{LSJ\tau} (C^{J\tau}_{A,B})^2 b_{S\tau}^2 \sum_M \left|\sum_N G^{NLS}_{J\tau} B^{M}_{NL}\right|^2 . \nonumber
\end{equation}

Here, $b_{S\tau}^2$ collects all spectroscopic factors and isospin coefficients of Table~\ref{Tab1} for the light ions. In this case is always $1/2$ as it absorbs an extra $(2S+1)^{-1}$ factor~\cite{Norman}. Notice also that $L=0$ reduces all summation to one unique term except that of $N$. $B^{M}_{NL}$ is an integral of the corresponding distorted waves, the wavefunction of the particle transferred and the interaction responsible of the transfer. This is so, as this calculation is a no-remnant calculation and can be understood as an approximation of the full second-order calculation shown here. Furthermore, this integral includes all the reaction kinematics part that we will assume to be identical in both $T=0$ and $T=1$ cases.

The remaining ingredient: $G^{NLS}_{J\tau}$ englobe all the structure information from the heavy ion. As we assume a single $j^2$ configuration, this factor can be retained proportional to:

\begin{equation}
    G^{N0S}_{J\tau} \propto (2j+1)\sqrt{2S+1}\begin{Bmatrix} \ell & \frac{1}{2} & j \\ \ell & \frac{1}{2} & j \\ 0 & S & J \end{Bmatrix} \langle 1 0 N 0; 0|n \ell n \ell;0 \rangle \nonumber
\end{equation}
where \{\} is a 9-$j$ coefficient and $\langle | \rangle$ is a Moshinsky-Talmi bracket, 
where $N = 2n +\ell$ due to selection rules so that it will also be unique. Notice that this Moshinsky-Talmi bracket is the same for both $T=0$ and $T=1$ cases.

\begin{figure}
\centering\includegraphics[trim=60 190 60 160,clip,width=8.5cm,angle=0]{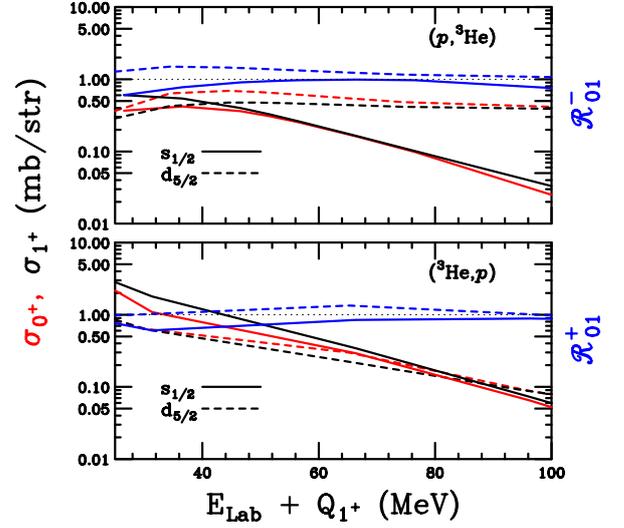}
\vspace{0.1cm}
\caption{(Color online) Left axis: Unit cross-sections for the $0^+$ (red) and $1^+$ (black) states and Right axis: $np$ Weisskopf Units (blue) as a function of the beam energy plus the Q-value to the $1^+$ state. Solid (dashed) lines correspond to $^{28}$Si ($^{20}$Ne) for  $(^3$He,$p)$  and  $^{32}$\replaced{S}{Si} ($^{20}$Ne) for $(p,^3$He).}
\label{fig3}
\end{figure}

Finally, taking into account all this factors, one can arrive to the result that:
\begin{equation}
\mathcal{R}_{01,2sp}^{+} \approx \mathcal{R}_{01,2sp}^{-} \approx \frac{1}{9} \frac{\begin{Bmatrix} \ell & 1/2 & j \\ \ell & 1/2 & j \\ 0 & 0 & 0 \end{Bmatrix}^2}{ \begin{Bmatrix} \ell & 1/2 & j \\ \ell & 1/2 & j \\ 0 & 1 & 1 \end{Bmatrix}^2          }
\label{eq8}
\end{equation}
the estimates from this approximation are also shown in Fig.~\ref{fig2}, to compare with those from Eq.~(\ref{eq6}). \added{Approximated ratios find an overall agreement with calculated ones. It could be noted that $\mathcal{R}_{01,2sp}^{+}$ and $\mathcal{R}_{01,2sp}^{-}$, which are approximately equal, are different due to kinematical aspects not included in the approximation.}  Furthermore, an inspection of Fig.~\ref{fig3}, where we present cross-sections and $np$ WU's as a function of the bombarding energy plus the Q-value to the $1^+$ state, for the representative cases of the $s_{1/2}$ and $d_{5/2}$ orbits, confirms that the ratios are stable even when the cross-sections change by factors of 10-100, and thus reflect a measure of the structural properties.

\section{Summary}
 Inspired by the works of Refs.~\cite{WU,Broglia72} we have introduced the concept of $np$ two-particle units (or $np$-Weisskopf units) to empirically assess, from the measured cross-sections and/or ratios, enhancement effects due to $np$-pairing correlations and the competition between the isoscalar and isovector channels.  
 
 Unit cross-sections and ratios were presented 
for the conditions relevant to the expected filling of the different $(n,j,\ell$) orbits along the N=Z line, from $^{16}$O to $^{100}$Sn. We believe that these units, used in the analysis of the $(^3$He,$p)$ and $(p,^3$He) reactions, will provide a simple and robust measure of $np$ pairing collectivity,  much in the same way as the Weisskopf units for electromagnetic transitions.

\section*{Acknowledgments}
We would like to thank Dr. Piet Van Isacker for for enlightening discussions.
This work is based on the research supported in part  by the Spanish Ministerio de Ciencia, Innovación y Universidades and FEDER funds under project FIS2017-88410-P, by the Director, Office of Science, Office of Nuclear Physics, of the U.S. Department of Energy under Contract No. DE-AC02-05CH11231 (LBNL) and by the U.S. National Science Foundation (NSF) under Cooperative Agreement No. PHY-1565546. Y.A. acknowledges the support by the Spanish Ministerio de Economía y Competitividad through the Programmes “Ramón y Cajal” with the grant number RYC2019-028438-I. \added{This work has received financial support from Xunta de Galicia (Centro singular de investigación de Galicia accreditation 2019-2022), by European Union ERDF, and by the “María de Maeztu” Units of Excellence program MDM-2016-0692 and the Spanish Research State Agency.}

\end{document}